\documentstyle[preprint,revtex]{aps}
\draft\tightenlines

\begin{document}
\preprint{LBL-32873}
\begin{title}
Can a strongly interacting Higgs boson rescue $\rm SU(5)$?
\end{title}
\author{Matthew H. Austern}
\begin{instit}
Lawrence Berkeley Laboratory, University of California,
Berkeley, California, 94720 \\ and \\
Physics Department, University of California at Berkeley,
Berkeley, California, 94720
\end{instit}
\receipt{\today}

\begin{abstract}
Renormalization group analyses show that the three running gauge
coupling constants of the Standard Model do not become equal at any
energy scale.  These analyses have not included any effects of the
Higgs boson's self-interaction.  In this paper, I examine whether
these effects can modify this conclusion.
\end{abstract}


\narrowtext

Although the Standard Model, a gauge theory based on the gauge group
${\rm SU}(3) \times {\rm SU}(2) \times {\rm U}(1)$, has
experimentally proved to be very successful, one unattractive feature
is that this gauge group is not simple, {\it i.e.,} that the theory
has three independent coupling constants, which are conventionally
denoted $g_3$, $g_2$, and $g_1$.  One possible solution is that this
gauge group is merely the low-energy manifestation of a simple Lie
group, which is spontaneously broken to ${\rm SU}(3) \times {\rm
SU}(2) \times {\rm U}(1)$ at some high energy scale.  A generic
feature of these ``grand unified theories'' (GUTs) is that the three
coupling constants evaluated at this scale are numerically equal.

At the one-loop level, it is straightforward to calculate the
$\beta$-functions of the three Standard Model gauge coupling
constants; the well-known result~\cite{beta-functions} is
\begin{equation}        \label{beta-function}
\mu {d g_i \over d\mu} = - {b_i \over 16 \pi^2} g_i^3,
\end{equation}
where, for $N$ generations of fermions and $n_h$ doublets of Higgs
scalars,
\begin{eqnarray}
b_1 &=&            - {4\over3} N - {1\over10} n_h        \\
b_2 &=& {22\over3} - {4\over3} N - {1\over6} n_h        \\
b_3 &=& 11         - {4\over3} N.
\end{eqnarray}
Integrating these equations,
\begin{equation}        \label{running-couplings}
\alpha_i^{-1}(\mu) =
  {b_i \over 2\pi} \ln \left({\mu\over\mu_0}\right) +
  \alpha_i^{-1}(\mu_0),
\end{equation}
where $\mu_0$ is some scale at which the coupling constants are
known.

The initial renormalization point, $\mu_0$, is arbitrary; it is
convenient to choose it to be $M_Z$.  The strong coupling constant,
$\alpha_3(M_Z)$, can be extracted from a variety of experiments; the
Particle Data Group~\cite{PDG} reports the average $\alpha_3(M_Z) =
0.1134 \pm 0.0035$.  The electroweak coupling constants,
$\alpha_1(M_Z)$ and $\alpha_2(M_Z)$, are obtained by the LEP
measurements of $\alpha_{EM}(M_Z)$ and $\sin^2 \theta_{\overline{{\rm
MS}}}$, yielding~\cite{SU5-study}
\begin{eqnarray}
\alpha_1(M_Z) &=& 0.016887 \pm 0.000040         \\
\alpha_2(M_Z) &=& 0.03322  \pm 0.00025.
\end{eqnarray}

Note that $\alpha_1$ is normalized in the way that is conventional
for discussions of GUTs, not in the way that is conventional for
discussions of the Standard Model.  That is,
\begin{equation}
\alpha_1(M_Z) =
       {5\over3}
       {\alpha_{EM}(M_Z) \over \cos^2 \theta_{\overline{{\rm MS}}}}.
\end{equation}
A different choice of normalization would require both a different
$\beta$-function and different matching conditions at the GUT scale.

Using the central values of these coupling constants, and assuming
three generations of fermions and a single Higgs doublet,
Eq.~(\ref{running-couplings}) is plotted in Fig.~\ref{GUT-triangle}.
Note that the three couplings do not meet at a single point;
unification is excluded~\cite{SU5-study} by more than $7$ standard
deviations.

A more careful analysis~\cite{Hall-paper,Jones-paper} would include
two-loop effects and matching conditions, but these effects turn out
to be small corrections~\cite{RG-study}, and the essential result is
unchanged.  A GUT is thus only possible if an intermediate
unification scale exists, or if additional physics, such as
supersymmetry, enters into the $\beta$-function,
Eq.~(\ref{beta-function}).  The simplest GUT, $\rm SU(5)$, possesses
no intermediate scale, so these results are usually assumed to
exclude non-supersymmetric $\rm SU(5)$ with the minimal scalar
sector.

In the Standard Model, it is well known that if the Higgs boson is
very massive, its self-interactions become strong.  None of the
renormalization-group analyses of the Standard Model include the
Higgs self-interaction, so it is interesting to consider whether this
might in fact be the additional physics that is necessary to achieve
unification.  This self-interaction cannot affect the one-loop
calculation of the gauge coupling $\beta$-functions, but, if it is
sufficiently strong, it might be significant even as a higher-order
effect.

Using the axial gauge, the renormalization of a gauge coupling
constant is related to the gauge boson's wave-function
renormalization by the Ward identity
\begin{equation}
Z_g = {1 \over \sqrt{Z_A}},
\end{equation}
so it suffices to consider vacuum polarization diagrams.  The number
of diagrams is further reduced by regarding ${\rm SU}(2) \times {\rm
U}(1)$ as a theory with unbroken gauge symmetry.  At an energy scale
$\mu$, any effect that depends on symmetry breaking will be
suppressed by some power of $v/\mu$, where $v \approx 246\ {\rm GeV}$
is the vacuum expectation value of the Higgs field.  As we are
dealing with scales up to $10^{16}\ {\rm GeV}$, this suppression
factor renders such effects completely negligible.

In the Standard Model, the Higgs scalars do not couple to gluons.
The lowest-order diagrams that might affect the $\rm SU(2)$ or $\rm
U(1)$ $\beta$-functions are given in Fig.~\ref{two-loop}.  In fact,
however, neither of these diagrams contributes to the wave-function
renormalization: Such contributions are only from divergent terms
that are momentum-dependent and that have the Lorentz structure
$g^{\mu\nu}$, and it is clear that neither of these diagrams has such
a form.  The lowest-order nonvanishing contributions are three-loop
diagrams, which are shown in Fig.~\ref{three-loop}.  These diagrams
have been evaluated by Curtright~\cite{Curtright} and
Jones~\cite{Jones-three-loop}.

In Curtright's notation, the scalar self-interaction is given by
\begin{equation}
{\cal L}_q = - {\lambda \over 4!} F_{ijkl}
               \phi_i \phi_j \phi_k \phi_l,
\end{equation}
where $\phi$ is a multiplet of real scalar fields.  Its covariant
derivative is given by
\begin{equation}
\left({\cal D}_\mu \phi \right) _i
 =   \partial_\mu \phi_i
   + i g A^a_\mu \left(T^a \right)_{ij} \phi_j,
\end{equation}
where $T^a$ is a representation matrix of the gauge group.  The
contribution of ${\cal L}_q$ to the $\beta$-function, then, is
\begin{equation}        \label{quartic-beta}
\beta_q = {1 \over 4!}
          \left( {1 \over 16\pi^2} \right)^3
          g^3 \lambda^2 T F,
\end{equation}
where
\begin{equation}
T \equiv {\rm Tr} \, \left( T^a \right)^2
\end{equation}
and $F$ is defined by
\begin{equation}
F_{iklm} F_{jklm} \equiv F \delta_{ij}.
\end{equation}

In the minimal Standard Model with one Higgs doublet, the quartic
scalar self-interaction is given by
\begin{equation}
{\cal L}_q = \lambda \left( \left| \Phi \right|^2 \right)^2,
\end{equation}
where the relation between $\lambda$ and the mass of the physical
Higgs boson is
\begin{equation}
m_H = v \sqrt{2\lambda(v)}.
\end{equation}
For this model, $F = 72$, $T_{\rm SU(2)} = 1$, and
$T_{\rm U(1)} = 25/9$.

The contributions to the $\rm SU(2)$ and {\rm U(1)}
$\beta$-functions, then, are
\begin{equation}        \label{quartic-beta-SM}
\beta_q = {24 \pi \over \left( 16 \pi^2 \right)^3}
          \alpha_i^2 \lambda^2 T_i.
\end{equation}

This contribution to the running coupling may be calculated by
rewriting the renormalization group equation as an integral equation,
and solving it by iteration.  The one-loop result has been given in
Eq.~\ref{running-couplings}, and the correction from
Eq.~\ref{quartic-beta-SM} is given by
\begin{eqnarray}
\delta \alpha_i(\mu) &=&
  {24 \pi T_i \over \left( 16 \pi^2 \right)^3}
  \int_0^{\mu/\mu_0} dt \,
        \alpha^2_i (\mu_0 e^t) \lambda^2(\mu_0 e^t)             \\
&\approx&                       \label{iterative-shift}
  {24 \pi T_i \over \left( 16 \pi^2 \right)^3}
  \int_0^{\mu/\mu_0} d t
  {\lambda^2(\mu_0 e^t) \over
    \left(   {b_i \over 2\pi} t
          +  \alpha_i^{-1}(\mu_0)
    \right)^2
  }.
\end{eqnarray}
If the running of $\lambda$ can be neglected, {\it i.e.,} if
$\lambda(\mu) \approx \lambda(\mu_0)$, this integral can easily be
evaluated, yielding
\begin{equation}
\delta \alpha_i(\mu) =
  {48 \pi^2 \lambda^2 T_i \over \left( 16 \pi^2 \right)^3}
  {\alpha_i^2(\mu_0) \ln \left({\mu\over\mu_0}\right)
                \over
   2\pi + b_i \alpha_i(\mu_0) \ln \left({\mu\over\mu_0}\right)
  }.
\end{equation}

If $\lambda$ is taken to be the largest value that could possibly be
reasonable, {\it i.e.}, $\lambda = 4\pi$, which corresponds to $m_H
\approx 1.2\ {\rm TeV}$, then
\begin{eqnarray}
\delta \alpha_1(10^{15}\ {\rm GeV}) &=& 1.1 \times 10^{-4},   \\
\delta \alpha_2(10^{15}\ {\rm GeV}) &=& 6.7 \times 10^{-5}.
\end{eqnarray}
This shift is too small to have a significant effect on the failure
of coupling constant unification.

In fact, even these numbers are unrealistically large, because the
running of $\lambda$ cannot be neglected.  In theories with
self-interacting scalar fields, the running scalar coupling constant
increases with the energy scale, eventually diverging at some finite
scale.  It appears, although it has not been proved, that if a scalar
field theory is to be valid for arbitrary high scales, the coupling
constant must vanish~\cite{triviality}.

If ``triviality'' is a real phenomenon and not merely an artifact of
computational schemes, then scalar field theories must be regarded as
effective field theories, valid only up to some cutoff $\Lambda$.
The larger $\Lambda$ is taken to be, the smaller are the permissable
values of the coupling constant $\lambda$, and as $\Lambda$ is taken
to $\infty$, $\lambda$ is driven to zero.

The $\beta$-function for $\lambda$ may be approximated by neglecting
the contribution from gauge boson and fermion loops; this is
reasonable, since the domain of interest is when $\lambda$ is large.
The one-loop result is
\begin{equation}
\mu {d \lambda \over d\mu} = {3 \over 2\pi^2} \lambda^2,
\end{equation}
and
\begin{equation}
\lambda(\mu) = {1 \over
                   \lambda^{-1}(\mu_0) -
                   {3\over 2\pi^2} \ln \left({\mu\over\mu_0}\right)}.
\end{equation}
Requiring that $\lambda$ be finite up to the cutoff $\Lambda$ yields
the upper bound
\begin{equation}
\lambda(\mu_0) \leq {2\pi^2 \over 3}
                    {1 \over
                     \ln \left({\Lambda\over\mu_0}\right)}.
\end{equation}
In particular, in the minimal $\rm SU(5)$ model, the cutoff $\Lambda$
must be at least the GUT scale.  Taking $\Lambda = 10^{16}\ {\rm
GeV}$ then implies $\lambda(v) \leq 0.21$, and $m_H \leq 160\ {\rm
GeV}$.  Experience with other techniques~\cite{triviality,RG-study}
suggests that this simple one-loop result is at least qualitatively
correct.

Taking the running of $\lambda$ into account,
Eq.~\ref{iterative-shift} becomes
\widetext
\begin{eqnarray}
\delta \alpha_i(\mu)
&=&
  {24\pi T_i \over \left(16\pi^2\right)^3}
  \int_0^{\mu/\mu_0}
  {dt \over
            \left(\alpha_i^{-1}(\mu_0) + {b_i \over 2\pi} t \right)^2
            \left( \lambda^{-1}(\mu_0) -
                   \left({3\over2\pi^2}\right)t
            \right)^2
  }                                                             \\
&=&
  {3 T_i \over (16\pi)^2}
  {\alpha_i^2(\mu_0) \lambda^2(\mu_0)
                \over
   3\lambda(\mu_0) + \alpha_i(\mu_0) b_i \pi}
  \Biggl[ {b_i \alpha_i^2(\mu_0) t \over
           2\pi + b_i \alpha_i(\mu_0) t}
        + {9\over\pi} {\lambda^2(\mu_0) t \over
                       2\pi^2 - 3\lambda(\mu_0) t}      \nonumber \\
                            &&
        \hskip 3.7cm                            
        + {6 b_i \alpha_i(\mu_0) \lambda(\mu_0)
                \over
           3\lambda(\mu_0) + \alpha_i(\mu_0) b_i \pi}
          \ln\left({2\pi^2 + \alpha_i(\mu_0) b_i \pi t
                        \over
                   2\pi^2 - 3\lambda(\mu_0) t } \right)
  \Biggr],
\end{eqnarray}
where $t = \ln (\mu/\mu_0)$.  Taking $\lambda(M_Z) = 0.2$ in
accordance with the ``triviality'' limit, $\delta\alpha$ is plotted
in Fig.~\ref{run-lambda-shift}.  Note that $\delta\alpha$ is
completely negligible until extremely close to the scale where
$\lambda$ diverges. At this scale, however, this calculation no
longer makes sense: By definition, this is the scale at which the
physics of the Higgs sector can no longer adequately be described by
a scalar field.
\narrowtext

The contribution from the Higgs self-interaction is too small to
affect the conclusion that the minimal $\rm SU(5)$ model is ruled
out.  This contribution can, however, potentially be as large as some
of the two-loop terms, and should be included in any precision
analyses.

\acknowledgments
I wish to thank Robert Cahn and Lawrence Hall for helpful
discussions.  This work was supported by the Director, Office of
Energy Research, Office of High Energy and Nuclear Physics, Division
of High Energy Physics of the U.S. Department of Energy under
Contract DE-AC03-76SF00098.

\figure{
\label{GUT-triangle}
One-loop calculation of the running of the $\rm SU(3)$, $\rm SU(2)$,
and $\rm U(1)$ coupling constants, assuming that the low-energy
particle content is a single complex scalar doublet and three
generations of quarks and leptons.  The three lines represent the
central values of the coupling constants, and the shaded regions
represent the one-$\sigma$ errors.  Note that although they approach
a similar order of magnitude at high energies, the three constants
never become equal.
}

\figure{
\label{two-loop}
Two-loop diagrams involving the four-point scalar self-interaction
term.  The external gauge bosons are those of ${\rm SU}(2) \times
{\rm U}(1)$.  Neither diagram contributes to the vacuum polarization,
nor, as a consequence, to the $\beta$-function.  }

\figure{
\label{three-loop}
Three-loop diagrams involving the four-point scalar self-interaction.
The external gauge bosons are those of ${\rm SU}(2) \times {\rm
U}(1)$.  These diagrams are the lowest-order contribution to the $\rm
SU(2)$ and $\rm U(1)$ gauge coupling $\beta$-functions that involves
that interaction.
}

\figure{
\label{run-lambda-shift}
Contribution to the running of $\alpha_2$, the $\rm SU(2)$ coupling
constant, from the Higgs self-interaction.  The quartic coupling
constant $\lambda$ has been taken to be 0.2, which is the maximum
possible value if the theory is to be valid up to $10^{16}\ {\rm
GeV}$, and is evolved according to the one-loop $\beta$-function.
Note that this contribution to $\alpha$ is negligible until the scale
where $\lambda$ diverges.  }

\end{document}